\documentclass[fleqn,usenatbib]{mnras}

\usepackage[T1]{fontenc}

\DeclareRobustCommand{\VAN}[3]{#2}
\let\VANthebibliography\thebibliography
\def\thebibliography{\DeclareRobustCommand{\VAN}[3]{##3}\VANthebibliography}

\usepackage{graphicx}   
\usepackage{amsmath}    
\usepackage{multirow}

\begin{document} 


\title[New Herbig-Haro flows in Mon~R2 south]{Complex investigations of an active star-formation region in southern part of Mon R2}

\author[T.A. Movsessian et al.]{
T.A. Movsessian,$^{1}$\thanks{E-mail: tigmov@web.am}
J. Bally$^{2}$\thanks{E-mail: john.bally@colorado.edu}
T.Yu. Magakian,$^{1}$\thanks{E-mail: tigmag@sci.am}
and A.V. Moiseev$^{3}$\thanks{E-mail: moisav@sao.ru}
\\
$^{1}$Byurakan Observatory NAS Armenia, Byurakan, Aragatsotn prov., 0213,Armenia\\
$^{2}$University of Colorado at Boulder,
Center for Astrophysics and Space Astronomy, Boulder, Colorado 80309-0389, USA\\
$^{3}$Special Astrophysical Observatory, N.Arkhyz, Karachaevo-Cherkesia, 369167 Russia\\
}

\date{Accepted XXX. Received YYY; in original form ZZZ}

\pubyear{2024}


\label{firstpage}
\pagerange{\pageref{firstpage}--\pageref{lastpage}}
\maketitle

\begin{abstract}

We continue to present the results of a Byurakan Narrow Band Imaging Survey (BNBIS). 
In this work we present the results of the search and further detailed investigation 
of the objects, found in the course of the BNBIS survey in the southern part of the Mon~R2 
association. For the search of HH objects the narrow band images, obtained with the 1-m Schmidt 
telescope of the Byurakan Observatory, were used. Newly found objects were imaged in optical 
and near-IR range with the Apache Point Observatory 3.5 meter telescope, and observed  
spectrally with long-slit spectrograph and scanning Fabry-Perot interferometer on 6 m telescope 
of Special Astrophysical Observatory of the Russian Academy of Sciences using SCORPIO-2. 
We found three new HH groups: HH~1233, HH~1234 and HH~1235, two of them represent extended 
collimated flows. HH~1233 is the C-shape bipolar outflow system associated with the 
2MASS 06084223$-$0657385 source star. HH~1234 is the helical chain of HH knots near the 
star V963~Mon. HH~1235 is a separate compact knot, connected with the visible only in 
mid- and far-IR source WISE J060856.57$-$070103.5. We found also several molecular 
hydrogen outflows, one of which coincides with HH~1233 and two other 
are associated with the deeply embedded IR sources in the same field. 
One more probable bipolar H$_2$ outflow is related to  WISE J060856.57$-$070103.5. 
The emission spectra and spectral energy distributions of the source stars were analyzed. 
According to them they should be under rather early evolutional stage.  

\end{abstract}

\begin{keywords}
Open clusters and associations: individual: Mon R2 South; Herbig-Haro objects; ISM: jets and outflows; Stars: pre-main sequence
\end{keywords}



\section{Introduction}
More than a half-century ago, the term `R association'  was used to describe compact groups of visual-wavelength reflection nebulae produced by stars embedded in interstellar dust clouds.    Although this term has disappeared in the recent literature,   R associations  continue to be of interest as tracers of  embedded low-mass  star formation.   R associations contain reflection nebulae of various sizes which are often illuminated  by young stellar objects.   Moreover, they often contain bright infrared sources which trace the young stellar population.      

The Mon~R2 association is a large elongated cluster of reflection nebulae, divided in several compact groups. Mon~R2 was included in the first list of R-associations by \citet{VandenBergh} and studied in more detail by \citet{HR}.  In addition to late B-stars which illuminate reflection nebulae, it contains a number of lower mass pre-main-sequence objects \citep{HerbigBell}.
Evidence of ongoing active star-formation was provided by the discovery of   several Herbig-Haro (HH) objects and molecular hydrogen objects (MHOs) located in this association \citep{CarballoEiroa, wang2005, Hodapp2007}.
For a review of the Mon R2 region and a list of its stellar content see  \citet{CH}. 

Mon~R2 is located in a large molecular cloud,  L~1646,  at a distance of 830$\pm$50 pc \citep{HR,CH}.   The mass of this cloud, which extends over a 3\degr $\times$6\degr\ (44 by 88 pc) region has been estimated  to be 9$\times$10$^4$ M\sun\ \citep{Maddalena}. The most active site of star formation is in the Mon~R2 main core,  located in the western part of the cloud, which has spawned a cluster of at least 370 stars.   A second cluster  centered on GGD~12-15  about 45\arcmin\ east of the Mon R2 main core has spawned at least 130 stars.   There are several other, smaller clusters and a distributed population of between 190 and 790 stars \citep{CH}.   

Our attention was focused on the the region located about 30\arcmin\ south from the Mon~R2 core which contains several nebulous stars, including V899~Mon, an eruptive object with spectral features of both FUors and EXors \citep{Ninan2015, Park2021}.   This region contains a compact group of infrared sources associated with small reflection nebulae. 

This study was initiated with observations taken as part of the Byurakan Narrow Band Imaging Survey (BNBIS) performed on 1~meter Schmidt telescope of Byurakan Observatory.   The goal was the discovery of new HH objects and collimated outflows from YSOs \citep{Movsessian2020}. Newly discovered objects were studied in more detail with near-infrared imaging and visual-wavelength, long-slit and integral field  spectroscopy. 
 
\begin{figure*}
\centering
    \includegraphics[width=210pt]{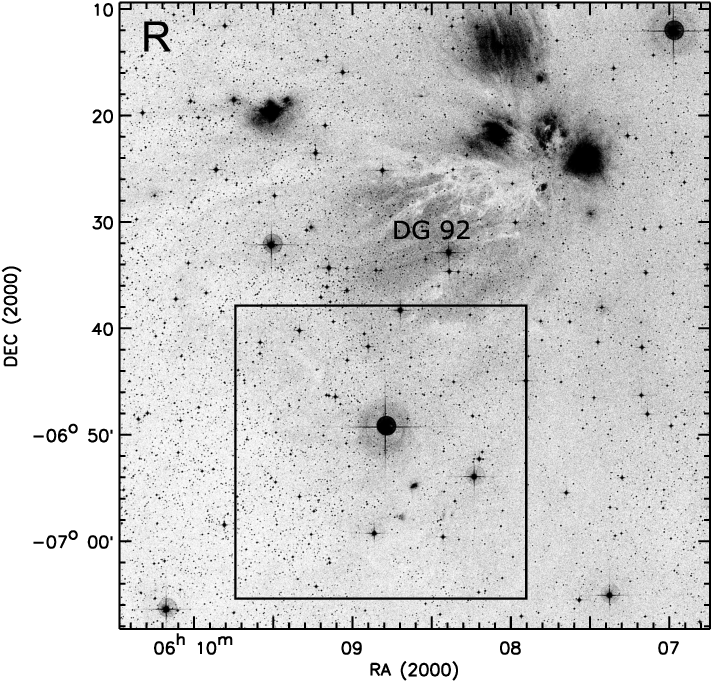}
    \includegraphics[width=220pt]{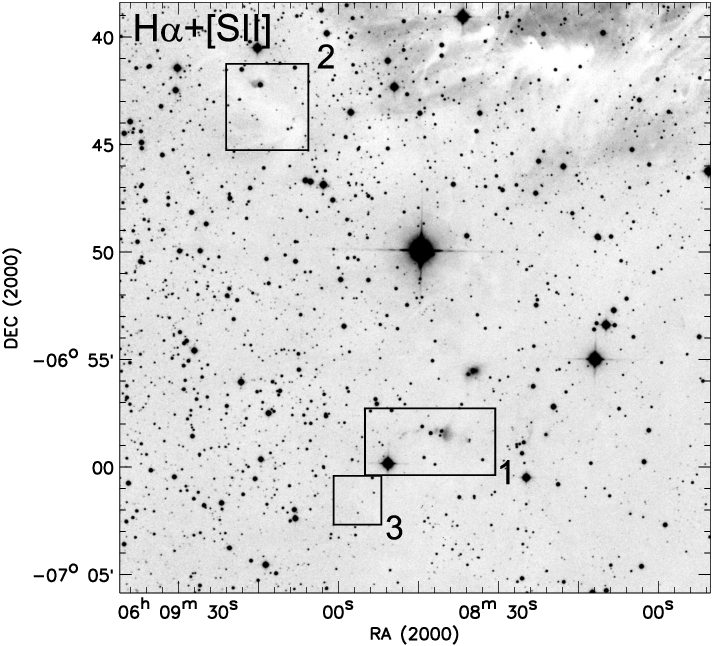}
\caption{Left: DSS-2 R image, showing the southern part of Mon R2 and the field observed in our study, reflection nebula DG 92 is marked; right:  the studied field in H$\alpha$+[S \textsc{ii}] emission,   boxes show the locations of newly discovered HH objects and outflows: HH~1233 (1), HH~1234 (2) and HH~1235 (3).  }
\label{fig1}
\end{figure*} 

\section[]{Observations}

\subsection{Visual wavelengths} 

Images were obtained on the nights of 21-27 January 2020 with 1~m Schmidt telescope at the Byurakan Observatory, which was equipped with  a reworked 4K$\times$4K Apogee (USA) liquid-cooled CCD camera providing  a 1-degree field of view with resolution of 0.868\arcsec\  per pixel \citep{Dodo}.     Narrow-band filters centered on 6560\AA\  and 6760\AA, both with 
FWHM of 100 \AA\ band-passes were used to obtain H$\alpha$ and [\ion{S}{ii}] images, respectively.    A medium band filter centered on 7500\AA\ with a FWHM of 250\AA\  was used for continuum imaging. 

A dithered set 5 min exposures were obtained in each filter.   The total  exposure in H$\alpha$ was 12,300 seconds, in [S\textsc{ii}],  5,400  seconds,  and in the continuum 2400 seconds. Images were reduced in the standard manner using the IDL{\footnote{IDL presently is a trademark of NV5 Geospatial.}} package developed by S.~Dodonov (Special Astrophys. Obs., Russian Acad. Sci.) which includes bias subtraction, cosmic ray removal, and flat fielding using super flats. 

Visual-wavelength H$\alpha$ images of HH~1233 and HH~1234 were also acquired with the Apache Point Observatory (APO) 3.5 meter telescope located near Sunspot, New Mexico on 22 March 2021.    Images were taken  the 2048 by 2048 pixel  
ARCTIC CCD camera with a narrow-band filter having  a 30\AA\ band-pass 
centered at 6570\AA .   Because of the f/10 focal ratio, the CCD was binned 3 by 3 to yield a pixel-scale of 0.347 arcseconds per pixel  and a field of view of 7.9\arcmin .    Three 300 second exposures were obtained on each source.  Standard procedures were used for bias and dark current removal and flat-fielding  using twilight flats.    The three images were registered using field stars and median combined to remove cosmic rays.   The effective exposure of the final H$\alpha$ images is 900 seconds.

\subsection{Infrared}

Near-IR observations of HH~1233 and V899 Mon were obtained with the APO  3.5-m telescope equipped with the NICFPS near-infrared imager on 22 March 2021.   NICFPS uses a  Hawaii 1-RG 1024$\times$1024 HgCdTe sensor with a 0.273\arcsec/pixel scale and 4.58\arcmin\ square  field of view  and a wavelength response from 0.85 to 2.4 $\mu$m.    Near infrared  H$_2$  emission was imaged using a 0.4\% bandpass filter centered at 2.12~$\mu$m using 180 second exposures.    During each observation, a set of 5 dithered images with offsets of 20\arcsec\ were obtained both on-source and on an off-source sky position located 5\arcmin\ from the on-source position. A median-combined set of unregistered,  mode-subtracted sky frames were used to form a master sky-frame that was subtracted from each individual on-source image.    Field stars were used to align  the sky-subtracted, on-source frames, which were median-combined to produce the final images which have effective exposure times of 900 seconds each.

\subsection {Long-slit spectroscopy}

Spectral observations were performed in 31 Dec. 2021 and 17 Dec. 2022\textbf{} on 6 m telescope of Special Astrophysical Observatory (SAO) of the Russian Academy of Sciences using SCORPIO-2 (Spectral Camera with
Optical Reducer for Photometrical and Interferometrical Observations \citep{Afanasiev2017}  multi-mode focal reducer mounted in the primary focus of the telescope in the spectroscopic
mode.    The detector,  CCD261-84 which has a  2048$\times$4104 pixel format with
the 15$\times$15-$\mu$m pixels,  was used for these observations. For  these observations the 
VPHG1200@540 grism in the wavelength range of 3650-7300\AA\ was used. The spectral resolution was about 5.2\AA\ across the full range of wavelengths (a mean reciprocal dispersion was 0.89\AA/px) and a spatial scale along the slit was 0.4\arcsec/pxl.

\subsection {Fabry-Perot interferometry}

Scanning interferometry  was carried out at the prime focus of the
6 m RAS/SAO telescope  on 07 and 08 Dec. 2020 in good atmospheric conditions with  2-2.5\arcsec\ seeing.
The SCORPIO-2 multi-mode focal reducer and Fabry-Perot interferometer were placed in the collimated
beam \citep{Afanasiev2017}. The
capabilities of this device in the scanning FPI observational mode are presented by \citet{moisav,
moisav2015}.
The CCD array,  EEV 40-90 $2\times4.5$K format  was used as a detector.
For the further analysis we used the sum of both  observations which was subjected to 2$\times$2  binning, so for each spectral channel,  512 $\times$ 512 pixel
images were obtained. The field of view was 6.1\arcmin\ with a scale of 0.71\arcsec\ per pixel.

The scanning interferometer's etalon, ICOS FPI was operated in the 751st order of interference at the H$\alpha$ wavelength, providing a spectral resolution with a  FWHM $\approx$ 0.4\AA\ (or $\approx$20 km s$^{-1}$) for a range of $\Delta\lambda$=8.7\AA\ (or $\approx$390 km s$^{-1}$) free from order overlap.    We observed 40  spectral channels with a channel spacing of  $\Delta\lambda\approx$ 0.22\AA\ ($\approx$10 km s$^{-1}$).   An interference filter with FWHM$\approx$15\AA\ centered on the H$\alpha$ line was used for pre-monochromatization.
 
We reduced our interferometric observations using the software developed at the SAO \citep{moisav,moisav2008,moisav2015} and the ADHOC software package{\footnote{The ADHOC software package was developed by J. Boulestex (Marseilles Observatory) and is publicly available in the Internet.}}. After primary data reduction, subtraction of night-sky lines, and wavelength calibration, the observational material represents ``data cubes''. We applied optimal data filtering, which included Gaussian smoothing over the spectral coordinate with FWHM = 1.5 channels and spatial smoothing by two-dimensional Gaussian with FWHM = 2--3 pixels.

\section[]{Results}

\begin{table*}
\label{HHknots}
 \centering
 \begin{minipage}{140mm}
  \caption{The coordinates of HH objects and HH flows in the field.}
  \begin{tabular}{lllll@{}}
  \hline
   Name     &  RA(2000)    & Decl.(2000) & Dist.(\arcsec )  & Notes  \\
 \hline\\
  \textbf{2MASS 06084223-0657385 flow}\\
  Central source & 06 08 42.2 & $-$06  57 38 \\ 
  \textit{Eastern branch} \\
  HH 1233 I   & 06 08 44.0 & $-$06 57 38 & 26 & \\
  HH 1233 J   & 06 08 44.7 & $-$06 57 39 & 37 & \\
  HH 1233 A  & 06 08 46.7 & $-$06 57 40  & 71  &extended along the flow    \\
  HH 1233 B  & 06 08 48.9 & $-$06 57 48  & 106 & \multirow{2}{*}{$\Bigr\}$ two  knots in common envelope} \\
  HH 1233 C & 06 08 49.4 & $-$06 57 53   & 108 & \\
  HH 1233 D & 06 08 50.0 & $-$06 58 02   & 117 & stellar-like\\
  HH 1233 E & 06 08 52.1 & $-$06 58 19   & 152 & probable head of the flow\\
  \textit{Western branch} \\
  HH 1233 K  & 06 08 39.4 & $-$06 57 43 & 40 & faint knot, elongated along the flow\\
  HH 1233 F & 06 08 36.8 & $-$06 57 59 & 82  & consists of several compact knots and wide trail\\
  HH 1233 G & 06 08 31.9 & $-$06 59 14   & 181 & faint and diffuse spot\\ 
  HH 1233 H & 06 08 32.1 & $-$06 59 33   & 190 & stellar-like, probable head of the flow \\ 
    & \\ 
  \hline\\
  \textbf{V963 Mon flow}\\
  Central source & 06 09 13.7 & $-$06 43 55\\
  HH 1234 A    & 06 09 13.0 & $-$06 43 52   & 11 & compact knot with two filaments \\
  HH 1234 B & 06 09 12.7 & $-$06 43 25   & 32 & knot with diffuse edges and two filaments  \\
  HH 1234 C & 06 09 11.2 & $-$06 43 14   & 56  & faint elongated knot\\
  HH 1234 D & 06 09 10.8 & $-$06 43 20   & 56 & faint elongated knot\\
  HH 1234 E & 06 09 09.9 & $-$06 43 31   & 61 & elongated, probable head of the flow\\
 \hline\\

\textbf{WISE J060856.57$-$070103.5 flow}\\ 
HH 1235 & 06 08 56.4 & $-$07 01 06      & & stellar-like knot

\\

  \hline

\hline  
\end{tabular}
\end{minipage}
\end{table*}

\subsection{Overview}

Figure \ref{fig1} shows an H$\alpha$+[S \textsc{ii}] image of the field  under investigation in the Mon~R2 region obtained with the 1-m Schmidt telescope.   The regions containing  new HH outflows  are marked by three rectangles.    New HH flows originate from V963~Mon, a little-studied eruptive variable star near V899~Mon,  from  2MASS 06084223$-$0657385, associated with a bipolar reflection nebula, and from the mid-IR source WISE J060856.57$-$070103.5.    A  description of results for each object is given below.
 

\subsection{HH~1233 in the 2MASS 06084223$-$0657385  region}

2MASS 06084223$-$0657385  is located about 30\arcmin\  south from the center of the DG~92 reflection nebula \citep{CH} in an area of moderate extinction (see Fig.\ref{fig1}, left panel).     Several nebulous objects can be seen in a  5\arcmin\  region on the DSS-2 and PanSTARRS images.  2MASS 06084223$-$0657385 is in the center of a bipolar reflection nebula with a symmetry axis  oriented roughly east-west.   The western lobe of the nebula is a wide-angle cone.   The eastern lobe is much fainter and detached from the central star by dark lane. It is likely that the western lobe is inclined toward us while  the eastern lobe is pointed away.    Comparison of the DSS-2 and PanSTARRS images, taken decades apart,  show no obvious changes in the shape and brightness of the nebula.  

\begin{figure*}
\centering
\begin{tabular}{@{}cc@{}}
  \includegraphics[width=220pt]{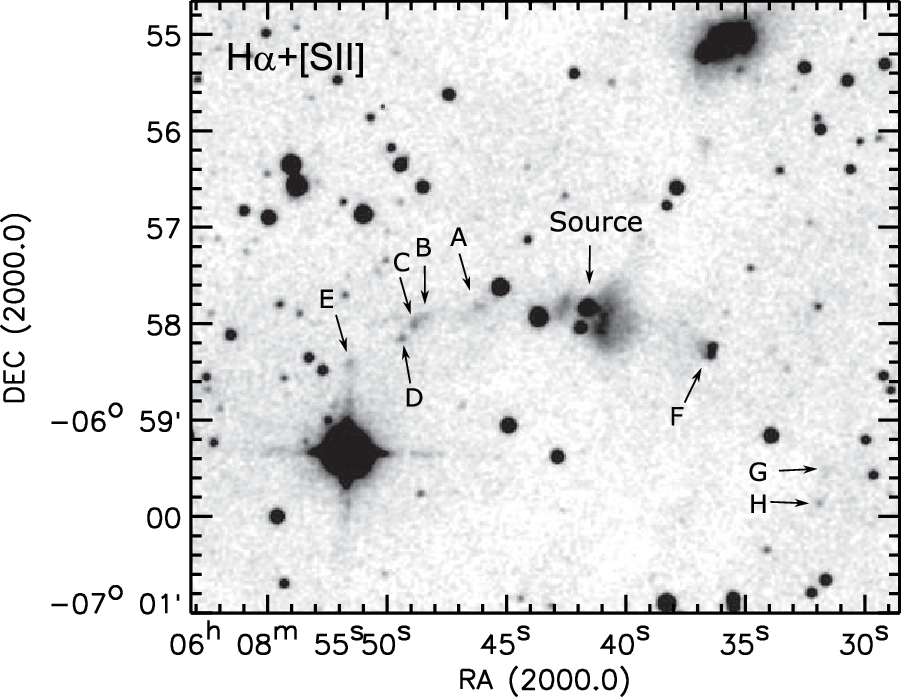}&
  \includegraphics[width=220pt]{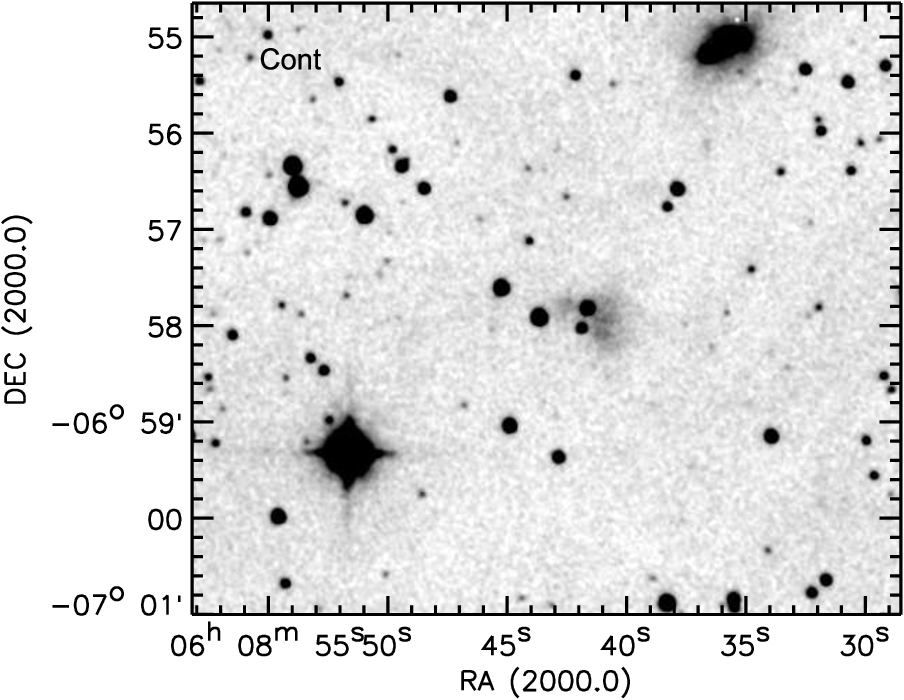}
\end{tabular}
\caption{HH 1233 flow near 2MASS 06084223$-$0657385.    H$\alpha$ + [\ion{S}{ii}] image (left panel) and continuum image (right panel) using data from the 1-m Byurakan Schmidt telescope.}
\label{HH1233}
\end{figure*}

\begin{figure*}
\centering
\begin{tabular}{@{}cc@{}}
  \includegraphics[width=220pt]{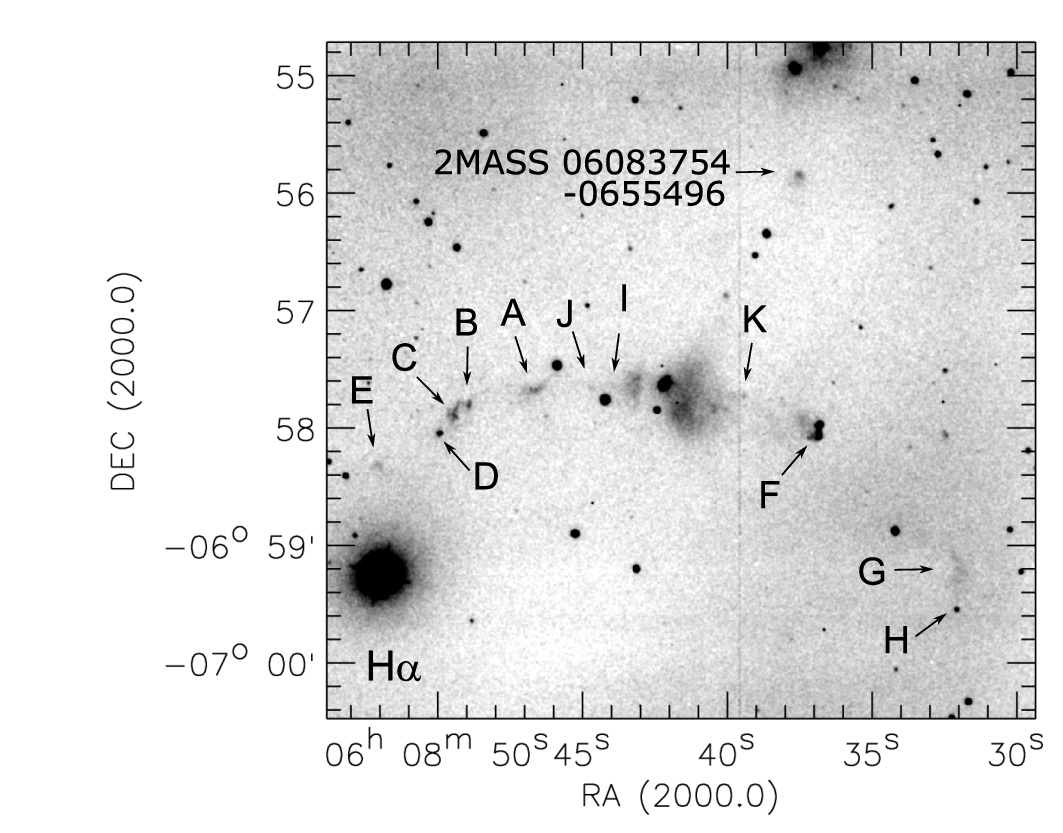} &
  \includegraphics[width=220pt]{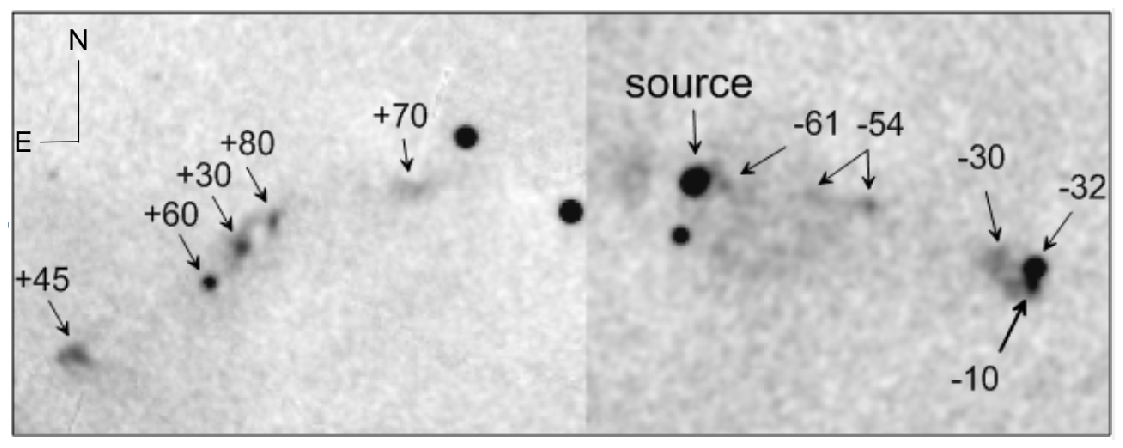}
\end{tabular}
\caption{HH 1233  observed  in  the H$\alpha$ line with the 3.5-m ARC telescope (left panel) and integrated FP image with the 6-m telescope (right panel). Radial velocities of the individual knots are indicated.}
\label{HH1233_HR}
\end{figure*}

The new images reveal a chain of HH knots near 2MASS 06084223$-$0657385 on  both H$\alpha$ and [\ion{S}{ii}] images. This HH flow is designated HH~1233.   The knots in HH~1233 traces a gentle, parabolic arc (see Fig. \ref{HH1233}, left panel).    Apparently, HH~1233 belongs to the class of C-symmetric flows which may indicate motion of the star, flow of the interstellar medium with respect to the star, or deflection of the flow by pressure or density gradients in the surrounding cloud.      For the more detailed analysis we used higher resolution images in the H$\alpha$ line, obtained with ARC telescope and the restored from FP observations  on 6~m telescope (Fig.\ref{HH1233_HR}).

The eastern branch of the outflow contains at least 5 bright separate knots, labeled as A-E.   All  are much brighter in H$\alpha$ than [\ion{S}{ii}]. Knot D is star-like.  The other knots have compact cores surrounded by diffuse envelopes.   As shown in Fig.\ref{HH1233_HR}, knots B and C are connected by a filament of emission.   However, their radial velocities differ significantly. The leading knot E at the eastern end  has the appearance of a pair of bow-shocks on the high-resolution FP image.

The shape of the western lobe mirrors the eastern lobe.  The western lobe  consists only of 3 bright knots: F-H (Fig.\ref{HH1233}). Knot F is the brightest in the whole flow and, unlike other knots, is  bright in [\ion{S}{ii}] emission. It has complex structure.   High-resolution images show at least three star-like clumps at its west edge with a diffuse tail extending east towards the suspected source.    A  faint star  is located near the northern edge of these clumps.   On the PanSTARRS images  Knot G  is a faint, diffuse feature.  Knot H marks the end of the western lobe of the flow; it is compact with slightly arcuate shape.        

The coordinates of the knots are given in Table 1. 
At a distance of 800 pc, the projected  length of this outflow is about 1.3 pc.  Thus, HH~1233  belongs to the class of  giant, parsec-scale HH flows. The Table 1 lists also the projected distances of the individual knots from the source star, measured by straight line (though the jet is curved).

\begin{figure*}
\centering
 \includegraphics[width=140mm]{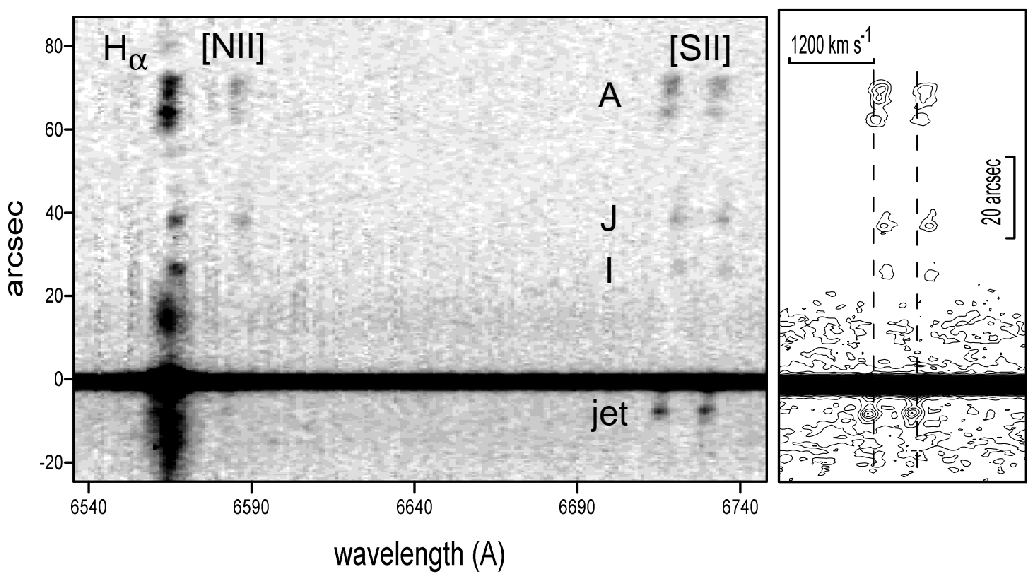} 
\caption{A part of the long-slit spectrum of 2MASS 06084223$-$0657385  near the H$\alpha$, [N\textsc{ii}] and [S\textsc{ii}] lines (left panel) and the PV diagram of [S\textsc{ii}] emissions (right panel). Zero radial velocities are shown by dashed lines. Positions of the jet and the knots HH 1233 A, I and J are marked. Note that the labeled knots are also visible in H$\alpha$ and [N\textsc{ii}] lines. Besides, the scattered H$\alpha$ emission is well visible near the star in the both lobes of the reflection nebula.    }
\label{LSdiag}
\end{figure*}

\begin{figure*}
\centering
\begin{tabular}{@{}cc@{}}
 \includegraphics[width=220pt]{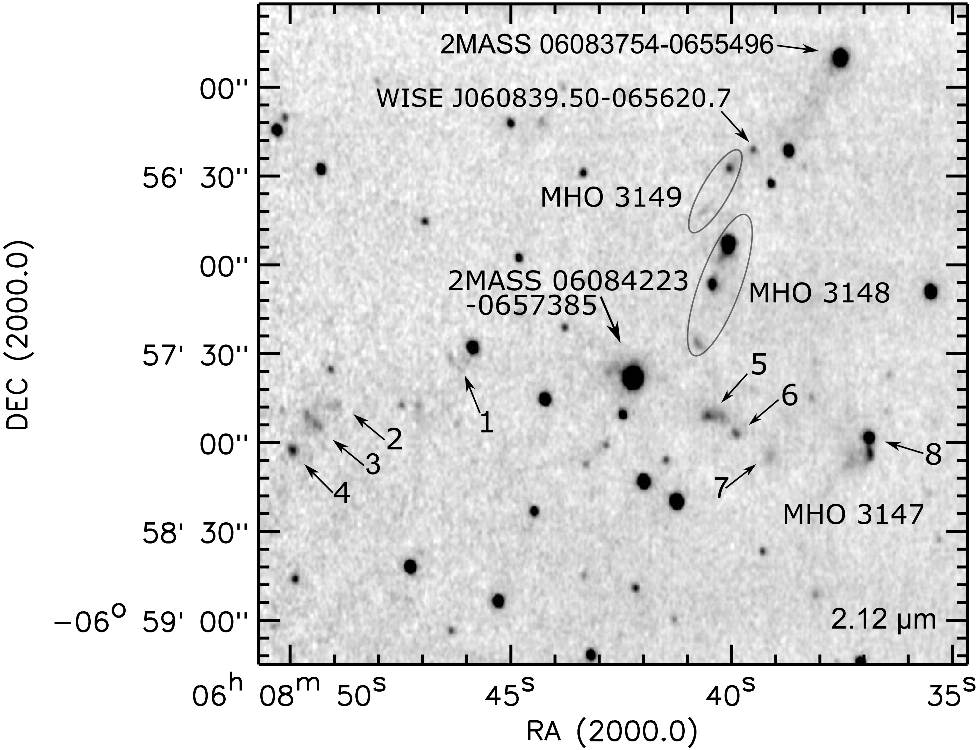}&
 \includegraphics[width=220pt]{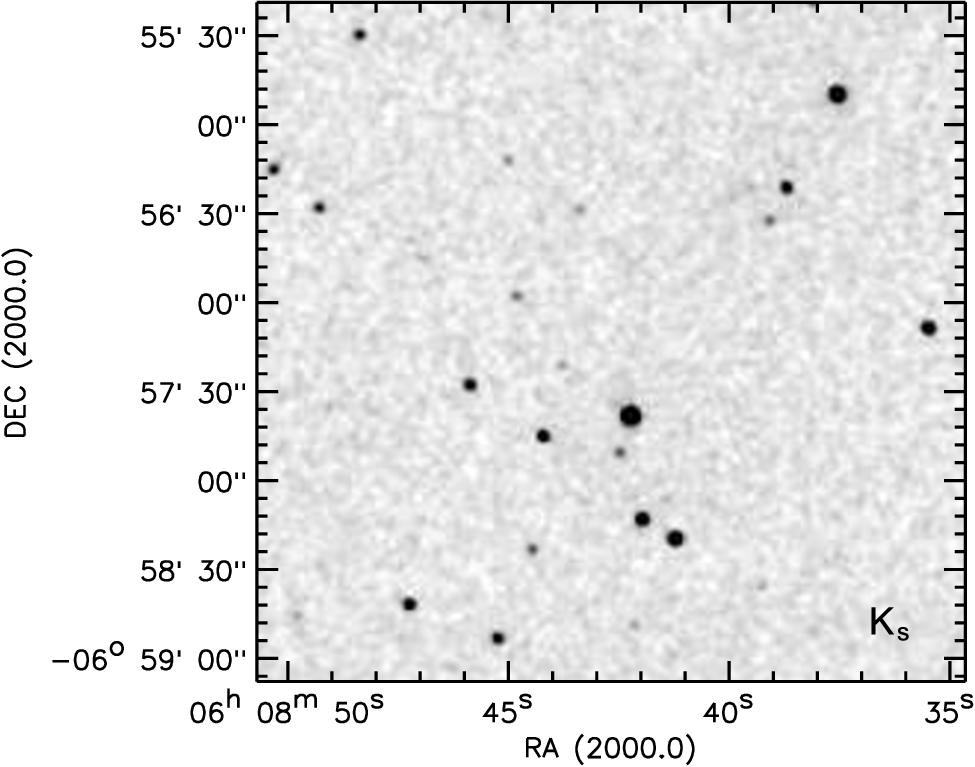}
\end{tabular}
\caption{Near-IR images of the 2MASS 06084223$-$0657385 region in 2.12 $\mu$m (H$_2$ emission), 3.5m ARC telescope (left panel) and in the 2MASS K$_{s}$ band (right panel). MHO 3148 and 3149 are marked by ellipses; the knots of MHO 3147 are shown by arrows and numbered.}
\label{HH1233_H2}
\end{figure*}

\begin{figure*}
\centering
 \includegraphics[width=300pt]{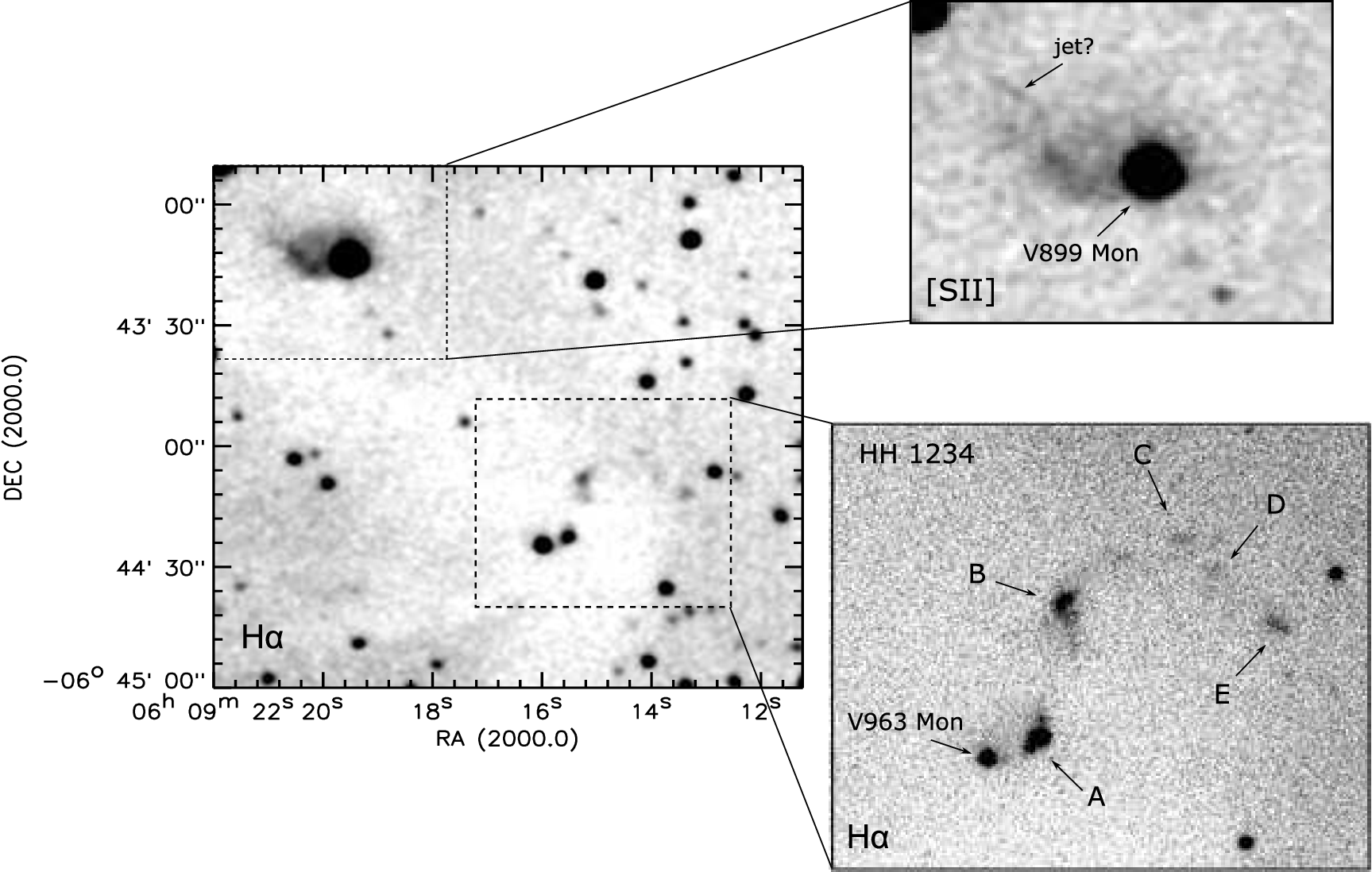}
\caption{Images of the probable jet near V899~Mon and HH~1234 near V963~Mon;   H$\alpha$, 1m Schmidt (left), enlarged [\ion{S}{ii}] image, 1m Schmidt (upper right) and 3.5m ARC telescope H$\alpha$ image (lower right). Individual knots of HH~1234 are marked by letters.}
\label{HH1234}
\end{figure*} 

\begin{figure*}
\centering
\includegraphics[angle=0,width=160mm]{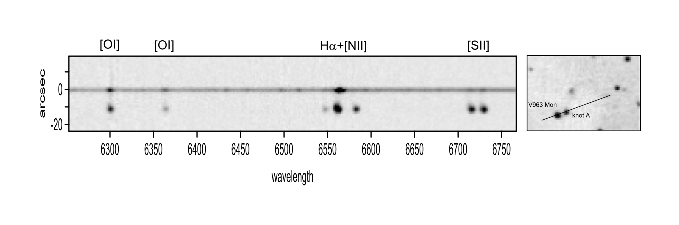} 
\caption{Spectrum of  V963~Mon star and of the knot HH~1234 A in red part. The large width of H$\alpha$ and metallic emissions in the stellar spectrum in the $\lambda\lambda$ 6400-6540 \AA\ are well seen. The right panel shows the orientation of the spectrograph slit on the direct image in  H$\alpha$ band; as can be seen, it misses other knots.}
\label{HH1234spec}
\end{figure*} 

\begin{figure*}
\centering
  \includegraphics[width=45mm]{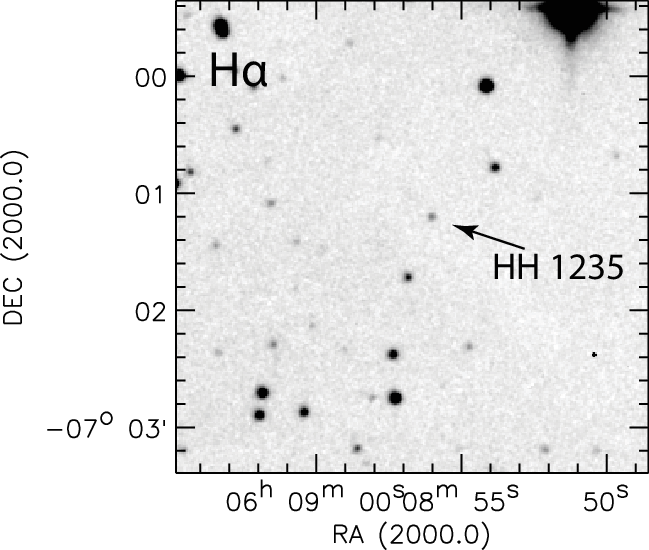}
  \includegraphics[width=42mm]{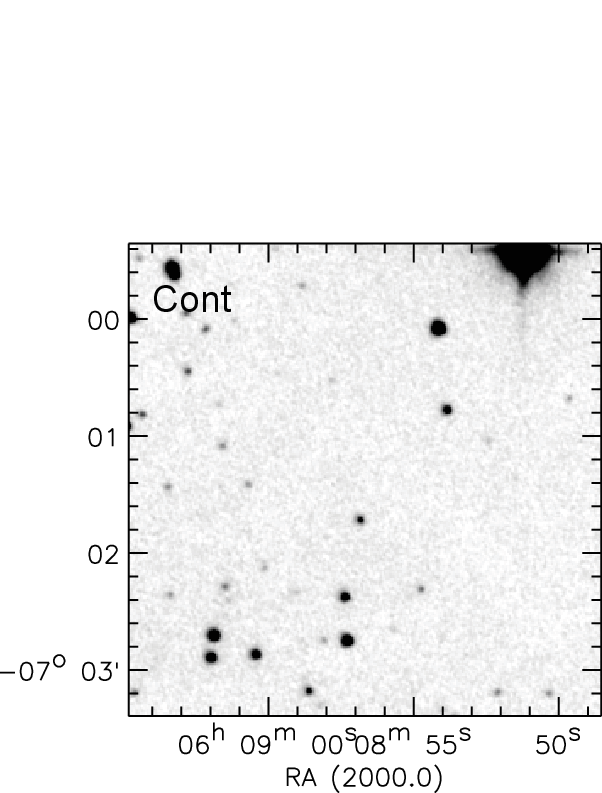}
  \includegraphics[width=43mm]{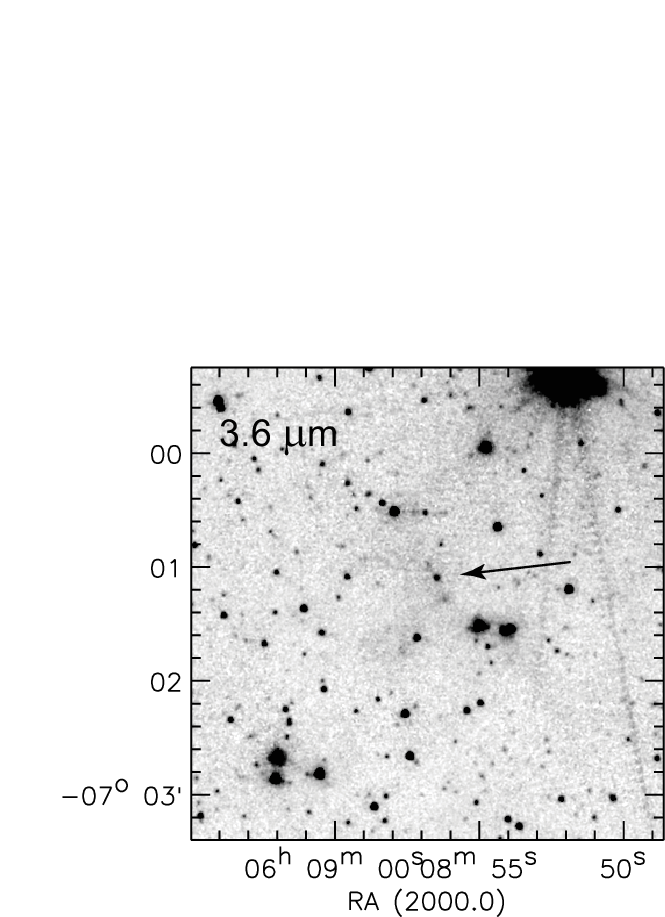}
  \includegraphics[width=42mm]{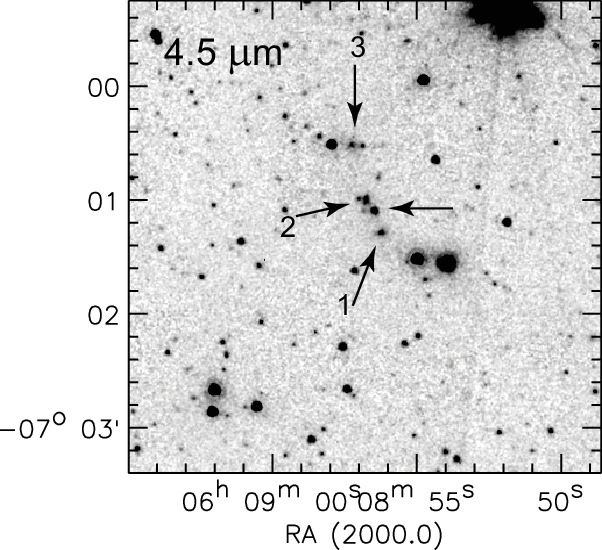}  
\caption{HH~1235 in H$\alpha$ and continuum, 1 m Schmidt images (two left panels); the same field in 3.6 $\mu$m and 4.5 $\mu$m bands of the Spitzer IRAC instrument (SEIP project) (two right panels). An arrow points to WISE J060856.57$-$070103.5 source, which coincides with HH~1235. The probable clumps of H$_2$ emission, bright  in 4.5 $\mu$m, are numbered. }
\label{HH1235}
\end{figure*} 

\begin{figure*}
\centering
\includegraphics{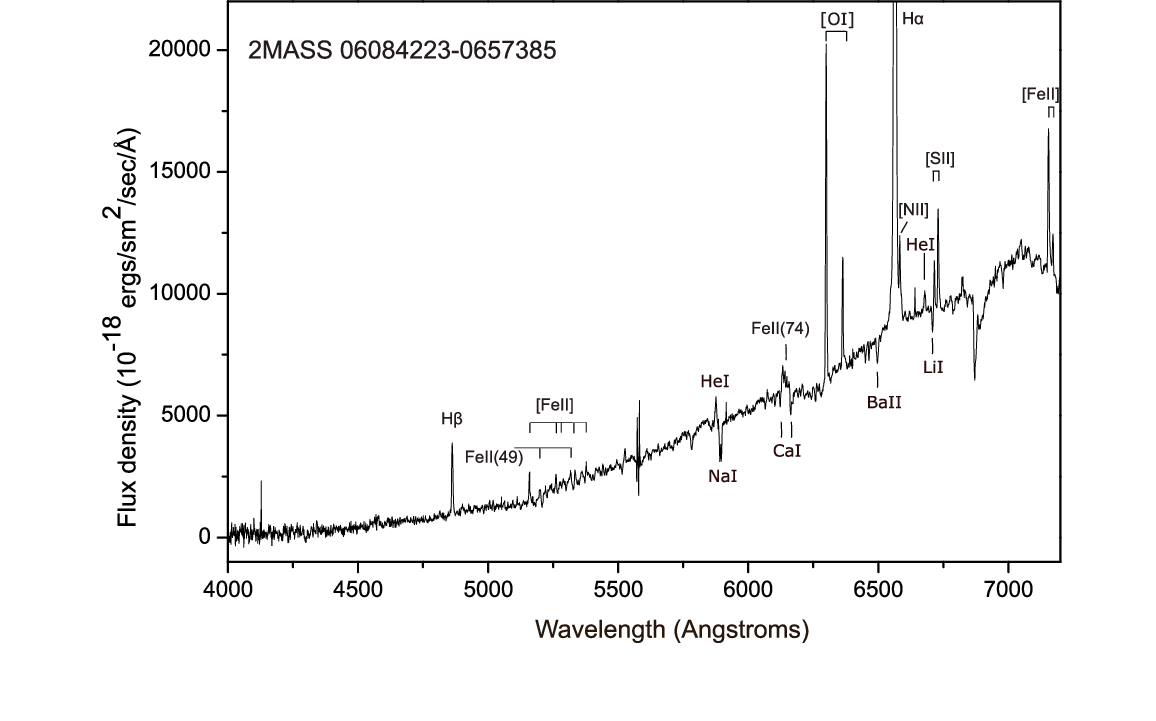} 
\caption{Spectrum of 2MASS 06084223$-$0657385 extracted from long-slit data.}
\label{2MASSspec}
\end{figure*} 

\begin{figure*}
\centering
 \includegraphics[width=150mm]{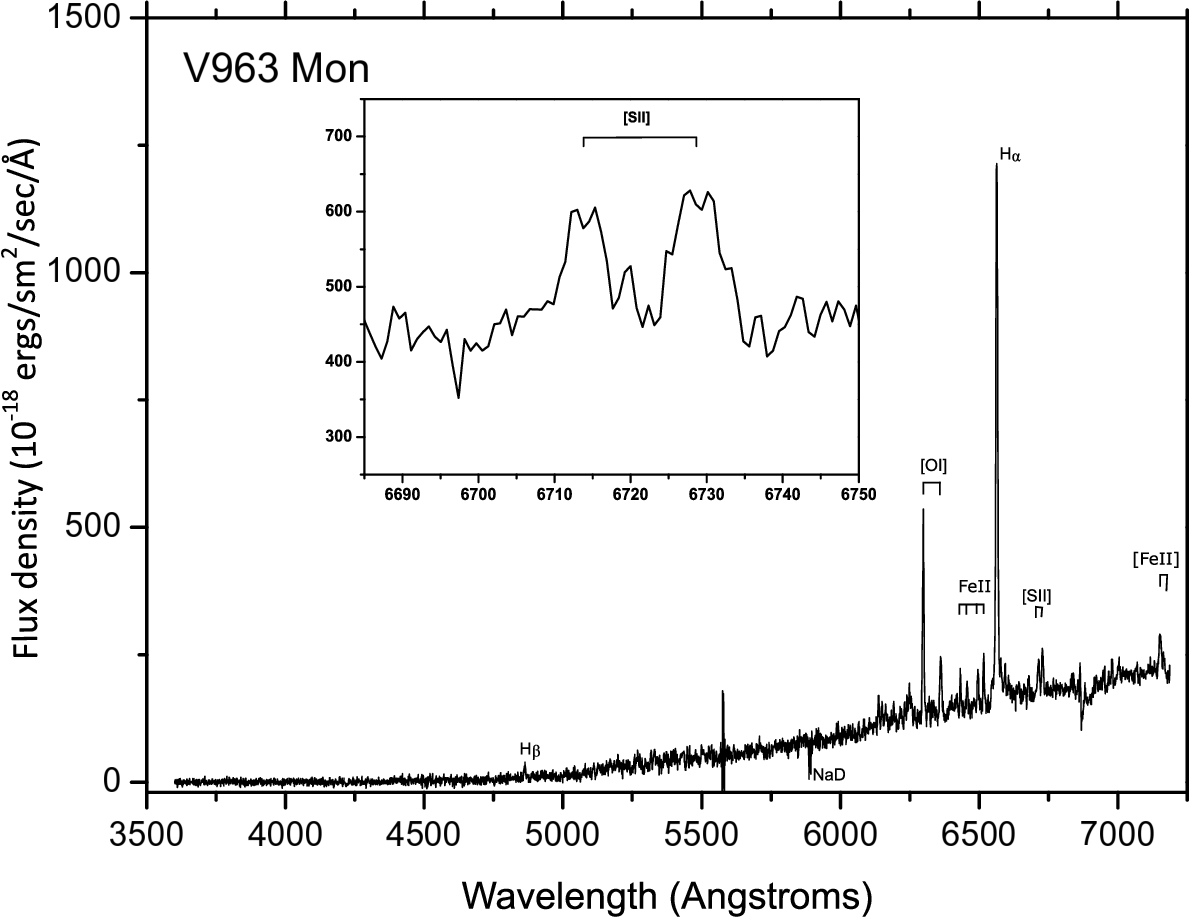} 
\caption{Spectrum of V963~Mon extracted from long-slit data. The inset shows an enlarged view of [SII] emission lines. Their split to the two or more components is obvious.}
\label{V963spec}
\end{figure*}  

We obtained a long-slit spectrum of the HH~1233 flow from  2MASS 06084223$-$0657385 using a  slit oriented in the E-W direction through the source and bipolar reflection nebula.   Figure \ref{LSdiag} (left panel) shows the part of two-dimensional spectrum and the position-velocity (PV) diagram  around H$\alpha$ and [SII] lines (right panel). The part above the spectrum of the star (discussed in the next section) corresponds to eastern direction. 

The stellar continuum and H$\alpha$ emission scattered by the reflection nebula can be seen toward lobes.
About 60\arcsec\ to east from the source the prominent emission lines of knot A are visible. Their extent is 14\arcsec, matching the visible size of the object on the images.  Next to H$\alpha$,  the lower intensity 6584\AA\  [\ion{N}{ii}] and 6716 and 6731\AA\ [\ion{S}{ii}] emission lines are visible.   The faintness of these lines explains why knot A is most visible on the  H$\alpha$ images.    Toward the source, two faint emission knots at distances of 27\arcsec\ and 37\arcsec\ from the star are seen.   These HH-knots are not visible on the 1~m images because they merge with an image of the nearby bright star. However, they are  discernible on the 3.5~m ARC telescope  H$\alpha$ image.   We labeled these knots I and J (Table 1).    Towards the west,  we found a blueshifted jet extending up to 10\arcsec\ from the central star.     A  compact,  bright knot is located   7\arcsec\ from the source.  This structure is not visible in direct images, likely because it is seen towards  the bright background of the reflection nebula.    

The PV diagram, built from long-slit spectrum (Fig.\ref{LSdiag}, right panel) shows the distribution of radial velocities in the eastern and western directions from the source. They  have the opposite signs:  the western lobe as whole has a negative  radial velocity,  while the eastern one has  positive radial velocity.    This confirms that the western lobe of the flow is inclined toward us.  There is  a large velocity gradient inside the knot A, which contains several condensations (see below).  Additionally,  the velocity dispersion between individual knots is large.   The intensity of the [\ion{N}{ii}]  and [\ion{S}{ii}] doublets increase from knot I to knot A.     This may either indicate decreasing extinction, or an increase in the excitation conditions with increasing distance from the source.     The 5007\AA\ line of [\ion{O}{iii}] only appears in  knot A (not shown), indicating higher excitation conditions than in the other knots. 

The flux ratio of the [\ion{S}{ii}]  lines provides an estimate of the electronic density.    Using the recently updated diagnostic diagrams \citep{POK}, we found that  the electron density in the jet is about 500 cm$^{-1}$, about 250 cm$^{-1}$ in  the knots I and J, while in the sub-condensations of the knot A (see next paragraph)  it varies from 100 to 190 cm$^{-1}$.   

Using this spectrum, we measured heliocentric radial velocities in the close to the star knots of HH~1233 using the emission lines of H$\alpha$, [\ion{S}{ii}],  and [\ion{N}{ii}].  However, close to the star,  the H$\alpha$ line could not be used  because it is heavily blended with the scattered stellar emission. Using the available emission lines ([\ion{S}{ii}],  [\ion{N}{ii}] and H$\alpha$, where possible),   we obtained the following mean values (in km s$^{-1}$): $-$95$\pm$7 (jet), +161$\pm$11 (knot I), +145$\pm$5 (knot J) and +30$\pm$9, +55$\pm$11 and +71$\pm$16 (three subsequent condensations from W to E  in knot A  seen on Fig.\ref{LSdiag}).

\begin{table*}
 \centering
 \begin{minipage}{160mm}
  \caption{ Newly found H$_{2}$ flows and their probable sources in the Mon~R2 South field.}
  \begin{tabular}{lcccccc@{}}
  \hline
   Flow  & Probable source   &  RA(2000)    & Decl.(2000) & Full projected & PA${^b}$ & comments \\
  & & h m s & \degr\ \arcmin\ \arcsec & length (pc)$^a$ & (degr) &  \\

 \hline
MHO 3147 & 2MASS 06084223$-$0657385 & 06 08 42.2 & $-$06  57 37 &  0.78 & $\approx$ 90 (curved) & 8 knots\\
MHO 3148 & 2MASS 06083754$-$0655496 & 06 08 37.6  & $-$06 55 50 & 0.34 & $\approx$150 & 3 knots or more \\
MHO 3149 & WISE J060839.50$-$065620.7 & 06 08 39.5 & $-$06 56 22 & 0.14 & $\approx$140 & at least 2 knots\\
& WISE J060856.57$-$070103.5 & 06 08 56.6  &  $-$07 01 04 & 0.19 & $\approx$200 & at least 3 knots\\

 \hline

\end{tabular}

\begin{flushleft}
\textit{Notes}: \\
$^{(a)}$  Determined for 830 pc distance. \\
$^{(b)}$ Position angle of the outflows measured from the probable source to east from north. \\

\end{flushleft}
\end{minipage}
\label{H2flows}
\end{table*}

To study the kinematics of the whole curved outflow  we used the data from the  scanning FP interferomer.    Figure \ref{HH1233_HR} (right panel)  shows the monochromatic H$\alpha$ image of the field around 2MASS 06084223$-$0657385 recovered from the FP data cube. Due to the smaller field of the FP mode this image does not cover the G and H knots at the west end of the outflow. This image shows both the positive and negative velocity channels of the FP data cube and has better  contrast than the narrow-band images. The jet on the west side of the source or rather an above described knot at its end can be clearly seen (marked by arrow with $-$61 km s$^{-1}$). The image also shows a previously undetected faint, elongated knot  near the edge of the conical reflection nebula.    This knot, labeled knot K,  is located between the source and  knot F (Fig.\ref{HH1233_HR}).     Faint filamentary structure connects several of the knots detected on the narrow-bans filter images. 

The heliocentric radial velocities of the knots determined from the FP data are shown in Figure  \ref{HH1233_HR}.    The radial velocity amplitudes exhibits large variations from knot to knot superimposed on a general trend of decreasing speed with increasing distance from the source.      The decreasing trend either indicates deceleration as ambient material is entrained or that the outflow is also curved along our line of sight  in addition to being bent on the plane of the sky.     The large knot-to-knot radial velocity variations suggest either large variations in the ejection velocity of the jet over time, or that the jet is interacting with a highly structured ambient medium.  
  
HH~1233 was imaged in the 2.12~$\mu$m emission line of H$_2$ with the APO  3.5-m  telescope.   Several new MHOs (molecular hydrogen objects) were found in this region.     The longest, numbered  MHO 3147, coincides with HH~1233 outflow.    Several knots in the bent HH~1233 flow coincide with H$_2$ condensations, namely B, C, D and F (see Fig.\ref{HH1233_H2}, left panel). At a projected distance of 1\arcmin\ northwest from the source of HH~1233,  we found two more molecular hydrogen flows flows (MHO 3148 and MHO 3149) which have no visual-wavelength counterparts. These flows may be powered by  2MASS 06083754$-$0655496 which is associated with a faint cone-shaped reflection nebula visible on the PanSTARRS i, z and y band images, and the deeply embedded source WISE J060839.50$-$065620.7. These two MHOs  consist of elongated chains of knots, whose axes pass through the suspected sources. These flows are invisible on the 2MASS K band image (see Fig.\ref{HH1233_H2}, right panel).    The MHOs are marked by ellipses in Fig.\ref{HH1233_H2} (left panel) along with the suspected sources, coordinates of which are listed in Table 2. Several small and very faint knots of H$_2$  emission can also be seen in the field which may trace shocks in additional outflows.

\subsection{HH~1234 in the V899 Mon region}

V899~Mon (IRAS 06068$-$0641) and V963~Mon (IRAS 06068$-$0643),   separated by  $\sim$2.5\arcmin\ from each other  are two recently discovered eruptive young stars  \citep{Wils2009}.  Although today V899~Mon is in a quiescent state, it underwent two FUor-like eruptions which lasted several years \citep{Ninan2015,Park2021}.     V899~Mon exhibits characteristics intermediate between FUors and EXors \citep{CR2018,Magakian}.      The star V963~Mon  varied between magnitude 15 and 20 over the years and exhibits deep fades reminiscent of  UX~Ori-type objects \citep{Wils2009}.      All published information about this little-studied star is contained  in the discovery paper \citep{Wils2009}.     Comparison of images taken by the DSS-2,  ZTF,  and PanSTARRS  surveys (available through Aladin Virtual observatory)  and our images shows that the  reflection nebulae, illuminated by both stars, are   variable.  

We obtained images of this field in H$\alpha$ and [S~\textsc{ii}] filters (Fig. \ref{HH1234}).     No emission knots were found near V899~Mon on the H$\alpha$ images.    However,  very faint, narrow,  jet-like feature is visible on the axis of the  V899~Mon reflection nebula in [S\textsc{ii}] emission (see Fig.\ref{HH1234} right top panel).  This object has a length about 11\arcsec\ and is located  about 42$\arcsec$ northeast from V899~Mon.     Deeper, high spatial resolution images are needed  to determine if this feature is indeed a jet from V899~Mon.  
If it really exists, this jet is  brighter in [S\textsc{ii}] than in H$\alpha$,  since 
on the H$\alpha$ images obtained with the APO telescope, nothing can be seen at its location.

A new HH flow, HH~1234, was found near V963~Mon.   V963~Mon is embedded in an  emission envelope.   The H$\alpha$ image shows a curving chain of HH knots extending about  60\arcsec\ northwest from V963~Mon.   The morphology resembles an inverted question mark.   The ARC 3.5-m  H$\alpha$ image reveals a faint emission filament  starting at  the western side of V963 Mon star which connects the brighter  HH knots (see Fig.\ref{HH1234} lower right panel). It is especially well seen between knots B and C.       The coordinates of the five knots are included in  Table 1. In [\ion{S}{ii}] this flow exhibits  a morphology similar to  H$\alpha$ though somewhat fainter.    

We obtained a long-slit spectrum of  V963~Mon with the slit oriented through the star and  knot A in  HH~1234 flow (see Fig.\ref{HH1234spec}). The spectrum of the knot A contains about fifteen emission lines, including  [\ion{O}{iii}], [\ion{N}{i}] and [\ion{Fe}{ii}] in addition to intense H$\alpha$, [\ion{O}{i}], [\ion{N}{ii}],  and [\ion{S}{ii}] , which indicates a wide  range of excitation conditions.   The bright emission lines such as H$\alpha$ are broad with a  FWHM of about  6-7\AA, implying a line width of  270 to 300  km~s$^{-1}$ . The heliocentric radial velocity of the HH 1234 A is $-$52 $\pm$ 6 km s$^{-1}$.      

The large line width, extreme curvature of  HH~1234, and the complete absence of a red-shifted counterflow suggests that the HH~1234 outflow axis may be  closely aligned with our line of sight. 

\subsection{HH 1235}

The HH~1235 stellar-like HH knot (Fig.\ref{HH1235}, left two panels) is located in the south-eastern direction from the HH~1233 flow (see Fig.\ref{fig1}, right panel).
It is visible  in H$\alpha$ as well as in the [\ion{S}{II}] emission lines. It is also visible on the PanSTARRS survey images.  About 5\arcsec\ to P.A.= 210\degr\ there is another, very faint emission knot. Nothing can be
seen in this field in 2MASS survey. However, a faint continuum source is seen in the AllWISE and unWISE surveys images. This source (WISE J060856.57$-$070103.5) is visible on the high quality 4.5 $\mu$m image of this area obtained from Spitzer SEIP project database. Fig.\ref{HH1235} (two right panels) shows that besides of this stellar source, at least three diffuse clumps are visible, located approximately in a straight line passing through the source under P.A. $\approx$ 200\degr.    They probably represent one more molecular hydrogen flow; thus we included it in Table 2.

\subsection{The Sources}

\subsubsection{2MASS 06084223$-$0657385}

According to the Gaia DR3 survey, the central star of HH~1233 flow is a close double with 0.24\arcsec separation and with the difference in brightness between components $\Delta$G=0.84 mag. More detailed information about it, including the distance, is not available.  We assume that the long-slit spectrum of the central source represents the brighter, southern component. 
Fig.\ref{2MASSspec} shows a part of the  spectrum of this star, extracted from the 2D image.
Comparing its appearance with red spectral standards \citep{AS}, we estimate its spectral type as G7-G9V by the strength and width of the absorption lines of \ion{Na}{i}, \ion{Ca}{i} and \ion{Ba}{ii}.    Strong absorption from  \ion{Li}{i}  with an equivalent width (EW) = 0.4 \AA\ is seen. 
 
The emission spectrum of this star is unusual.
The  Balmer series  of hydrogen emission lines are visible up to H$\gamma$ and possibly  H$\delta$.  In addition to  the shock-excited forbidden lines of [\ion{S}{ii}], [\ion{O}{i}], [\ion{N}{ii}], commonly seen in jets sources, we found prominent \ion{He}{i}  emission with an equivalent width of the $\lambda$ 5875\AA\ line of  1.6 \AA .   A large number of forbidden [\ion{Fe}{II}] lines are also present.     Surprisingly,   the permitted emission lines of  
\ion{Fe}{II} and \ion{Fe}{I} , commonly seen in T Tauri stars, are absent in the spectrum of this star.    The only permitted lines of \ion{Fe}{II} which we were able to reliably identify are several lines of the (49) and (74) multiplets.    The brightest  [\ion{Fe}{II}] lines are the $\lambda$\ 7154 \AA\ and $\lambda$\ 7170 \AA\  (multiplet 14F) which have EW $\approx$ 3.0 and 0.7 \AA\ respectively. 

The heliocentric radial velocity of Na D absorption is +26 km s$^{-1}$, of $\lambda\ 6122\ \AA\  \ion{Ca}{i}$ +2 km s$^{-1}$, of \ion{Li}{i} line +9 km s$^{-1}$, confirming their photospheric origin. The mean velocity of \ion{He}{i}  emission is +3 km s$^{-1}$.    The Balmer emission lines are slightly more redshifted: +39 km s$^{-1}$ (H$\alpha$);  +59 km s$^{-1}$ (H$\beta$). On the other hand, the mean velocity of  the five forbidden emissions lines of  [\ion{S}{ii}], [\ion{O}{i}], [\ion{N}{ii}] is $-47$ km s$^{-1}$, which is close to the velocity of the jet knots near the star (see Fig.\ref{HH1233_HR}). The velocity of the two strongest [\ion{Fe}{II}] lines ($-54$ km s$^{-1}$) also is very close to this value.     

\subsubsection{V963 Mon}

The low-resolution spectrum of V963~Mon, presented in the discovery paper of \cite{Wils2009}, does not allow  determination of  the spectral type of this star.     The only visible absorption is that due to the NaD doublet.    The number of emission lines are also low with nearly all in the near-infrared range such  as the very strong  \ion{Ca}{ii} triplet and two \ion{Fe}{i} lines. In the visible range  strong H$\alpha$ and several lines of ionized iron can be seen. 

Our spectrum of V963~Mon, extracted from the  long-slit data, is presented in the Fig.\ref{V963spec}.   The spectrum is similar to the one presented by  \citet{Wils2009}. The continuum is very red and virtually disappears below $\lambda$ 4800 \AA.   The NaD lines are  in absorption, but the low signal/noise ratio does not allow  analysis of their line profiles and velocities. No traceable molecular absorption bands were detected.    On the other hand, emission lines in the red range are intense.   Besides the single-peaked but wide  H$\alpha$,  faint H$\beta$,  and forbidden emission lines  of [\ion{S}{ii]}, [\ion{N}{ii}] and \ion{[O}{i}],   we identified several lines of \ion{Fe}{ii}, \ion{Ca}{i} and  two fluorescent lines of \ion{Fe}{i}. An unusual feature of V963~Mon spectrum is the  large width of the permitted emission lines (FWHM about 4-6 \AA), while the forbidden lines are even wider and obviously split into several components (at least [SII] lines, but probably also [OI]). This line split cannot be studied in detail due to the moderate spectral resolution of the data, but the difference between the centers and  the edges  of the forbidden lines is at least 150 km s$^{-1}$. The averaged radial velocity of forbidden lines is about $-$100 km s$^{-1}$, while for Balmer lines it is about +50 km s$^{-1}$.   

\subsubsection{WISE J060856.57$-$070103.5}

Very little is known about this object. It should be not confused  with the  brighter,  nearby IRAS
06064$-$0700, which is   located
in 52\arcsec\ to SW. Both sources are seen in WISE survey, as well as on the HERSCHEL PACS
images.   The coordinates
of the optical HH knot, mid-IR and far-IR sources do not coincide; they are shifted towards the  north
with respect  to each other by $\approx$\ 8\arcsec\ and PA about
40\degr\ (measured from HH~1235).
On the other hand, WISE J060856.57$-$070103.5 is embedded in a dark cloudlet $0.67\arcmin \times
0.33\arcmin$\ size, known as  the sub-mm source SCOPE G214.42$-$12.62 \citep{Eden}.

\section{Discussion and Conclusions}

The first question to be answered is whether all new sources of flows are really related to Mon~R2 complex. Gaia DR3 parallax for  V899~Mon (1.2358 $\pm$ 0.0261 mas) corresponds to a distance 809 $\pm$ 17 pc, which is close to the assumed 830 pc distance to Mon~R2.    However, the value of RUWE (renormalised unit weight error) is 1.99, which may be due to bright variable nebulosity near V899 Mon.    The probabilistic distance estimates, given in  \citet{BJ} catalog (778 and 785 pc),  seem less credible.

V963~Mon is probably embedded in the same cloud as V899 Mon. But Gaia DR3 data do not  support to this assumption.   The Gaia parallax for V963~Mon is  1.6651 $\pm$ 0.1376 mas (RUWE = 1.277), which implies a  significantly closer distance of 600  $\pm$ 50 pc. The estimated  errors from \citet{BJ} are nearly the same.    The influence of the nearby variable nebula probably affects the measurements.    

2MASS 06084223$-$0657385 was found by Gaia DR3 to be a close double star.   But neither its parallax nor proper motions were measured.  No Gaia astrometric data exist  for any of the nearby nebulous stars or the stars associated with new MHOs in this field. However, we assume that the small cloud  in which 2MASS 06084223$-$0657385 is located belongs to the one of filaments in Mon~R2 complex \citep{Kumar} at a distance of 830 pc.

\begin{figure}
\centering
  \includegraphics[width=50mm]{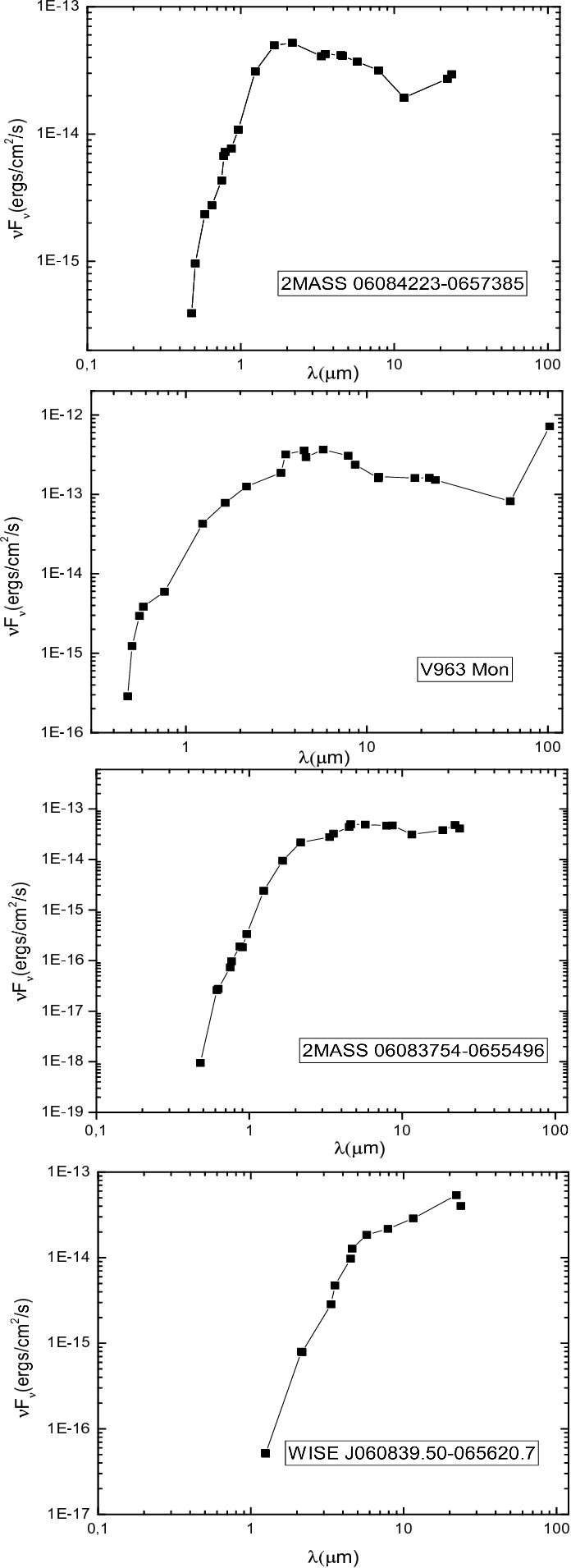} 
\caption{SEDs of the IR-sources.}
\label{SEDs}
\end{figure}

To better understand the nature of the newly found active stars in Mon R2 South, the spectral energy distributions (SEDs) were derived for  the outflow sources, using the available photometry  from Gaia, PanSTARRS, 2MASS, WISE and AKARI catalogs.  Figure \ref{SEDs}  shows SEDs for the two optical sources (2MASS 06084223$-$0657385 which is a close double, and V963~Mon) and for two sources of IR molecular flows (2MASS 06083754$-$0655496 and WISE J060839.50$-$065620.7).  The V899~Mon SED was presented  by \citet{Park2021}.         

2MASS 06083754$-$0655496  is not visible in the visual wavelength  range, and all its visual photometry refers to the small reflection nebula.    A far-IR IRAS measurement exists only for  V963~Mon which was included in  its SED.
No attempt was made to build SED  for WISE J060856.57$-$070103.5, since the very scarce amount of
photometric data; it is also unclear which data represent the real star -- probably only the far-IR ones (see sec. 3.5.3)

  First three SEDs  are similar and can be characterized as ``flat''.    WISE J060839.50$-$065620.7 is significantly redder than the other stars and may be a Class 1 YSO.    Lower limits  for  the bolometric luminosities for these sources were estimated from their SEDs.    Assuming a distance of 830 pc, we estimate L>2.5 L$_{\sun}$ for 2MASS 06084223$-$0657385, L>2.2 L$_{\sun}$ for 2MASS 06083754$-$0655496,  and L>1.3 L$_{\sun}$  for WISE J060839.50$-$065620.7.     V963 Mon is an order more luminous object with L>21~L$_{\sun}$ for assumed distance of  830 pc, and even if
its distance is 600 pc, its luminosity is L>11~L$_{\sun}$.
If the 100 $\mu$m IRAS measurement is excluded  from its SED because there may be other sources within the large IRAS beam which contaminate the flux estimate,   these luminosity estimates  decrease to  L>17 and 9 L$_{\sun}$,  respectively. 

The morphology of both HH~1233 and HH~1234 are unusual. The arcuate shape of HH~1233 suggests that it is a member of the subclass of C-symmetric flows in which the jet or outflow is deflected by a sidewind or the motion of the jet source through the medium.   Many externally irradiated jets located in HII regions such as the Orion Nebula or in soft-UV dominated environment such as NGC~1333 exhibit such C-shaped symmetry because of the flow of plasma or a stellar wind past the jet source  \citep{Bally&Reipurth2001}.     The environment  of Mon R2 is more similar to NGC 1333  region than the Orion Nebula because  to the lack of high-mass OB-stars  in the Mon R2 association.   

For the C-symmetric outflows in NGC 1333,  motion of a source relative to the interstellar medium was suggested \citep{Bally&Reipurth2001}.  But, unlike in  NGC 1333,   the HH~1233 curved outflow bends away from the core of the Mon R2 association.   It is possible  that the jet source has significant proper motion directed to the north, and the jet is deflected  by the interaction with the ambient medium. 
In this case PM vector should be directed to the center of Mon R2. Unfortunately, the proper motion
of the source star was not
measured by Gaia. (Can the binarity of 2MASS 06084223-0657385 play some role in the curvature
of flow?)

The helicoidal shape of HH~1234 is unique.   As seen from the source, various knots are spread over an  angle  of 41\degr .    A plausible explanation is that the circumstellar accretion disk responsible for determining the jet orientation is experiencing forced precession.       Most YSOs are born in dynamically unstable, non-hierarchical   multiple systems which tend to re-arrange into stable configurations consisting of compact binaries or hierarchical multiples plus ejected  stars \citep{Reipurth2010, Reipurth2012,Reipurth2015,Reipurth2023}.  Dynamical rearrangements often result in the formation of compact binaries in which the secondary orbit is eccentric and not in the same plane as the accretion disk surrounding the primary \citep{Moeckel2006,Cunningham2009}.  Such a configuration will drive forced precession of the disk. The precessing disk then creates a precessing jet.    If the jet is viewed nearly end-on, the precession angle will be greatly exaggerated.    We proposed that the source of HH~1234 is a close binary which drives a pulsed and precessing jet which oriented close to our line of sight.  Future proper motion and radial velocity measurements of the jet knots and photometric searches for periodic variability indicating a binary will test this hypothesis.

The narrow jet associated with  V899 Mon needs confirmation.    This star has photometric characteristics intermediate between FUOrs and EXors.   EXors to which this star is probably related, tend not to be associated with
collimated optical outflows.

The source stars are not  typical CTTS. The unusual spectrum of 2MASS 06084223$-$0657385 points to existence of an extended envelope, making it similar to ZZ Tau IRS, which represents a very extreme case \citep{Burlak}. V963 Mon demonstrates large amplitude brightness variations \citep[and references therein]{Wils2009} and has  wide and multicomponent spectral line profiles in  the forbidden emission lines, especially in [\ion{S}{ii}], as shown in the Fig.\ref{V963spec}. These characteristics are similar to the  EXor PV~Cep.

WISE J060856.57$-$070103.5  and its outflow may be  similar
to the well-known source and outflow, HH~30, where the stellar light is nearly completely absorbed by the seen edge-on
circumstellar disk.   We
see only the emission line jet and a small bipolar reflection nebula. In this case the far-IR source can
be associated  with the star itself, mid-IR emission is probably its reflected light, and  we observe the optical
ionized jet as HH~1235 along with a nearly symmetric H$_2$  flow.
This  may  be a  Class 0 young stellar object with a collimated outflow. 

Below we summarize our findings.

\begin{itemize}
\item 
We discovered a C-symmetric curved HH-outflow HH~1233, associated with the 2MASS 06084223$-$0657385. It is associated with the molecular hydrogen outflow MHO~3147.  
\item Long-slit spectroscopy confirmed the bipolar nature of this outflow,  reveal additional HH-knots,  and a  jet near the central source in the  bright reflection nebula.
\item A  narrow jet may exist on the axis of the reflection nebula associated with V899~Mon. 
\item A unique, helical chain of knots, designated HH~1234,  emerges from V963~Mon.   
\item NIR images of H$_2$ emission identify  two new molecular hydrogen  flows in the field containing MHO 3147. They are associated with 2MASS 06083754$-$0655496 and WISE J060839.50$-$065620.7.
\item The spectrum of 2MASS 06084223$-$0657385 is  rich with strong forbidden lines. Both permitted and forbidden emission lines in the spectrum of V963~Mon are  wide.   The  forbidden lines contain several components.
\item Our results confirm the ongoing star formation process in the southern edge of the Mon R2 complex.

\end{itemize}

\section*{Acknowledgments}

We thank the referee for helpful comments and suggestions. This work is a part of the Byurakan Narrow Band Imaging Survey (BNBIS). 
We obtained the SCORPIO-2 data on the unique scientific facility "Big Telescope Altazimuthal"
of SAO RAS as well as made data reduction with the financial support of grant No075-15-2022-262 (13.MNPMU.21.0003) of the Ministry of Science
and Higher Education of the Russian Federation. Some of the work presented here is based on observations obtained with the Apache Point Observatory 3.5-meter telescope, which is owned and operated by the Astrophysical Research Consortium.    We thank the Apache Point Observatory Observing Specialists for their assistance during the observations.  J.B. acknowledges support by National Science Foundation through grant No.  AST-1910393.
 This work was supported by the RA MES State Committee of Science, in the frames of the research
project number 21T-1C031.
We thank Prof. Bo Reipurth for providing the numbers for new HH objects and  for some helpful suggestions.
We are thankful to Dr. R. Uklein and D. Oparin for the FPI observations on the 6-m telescope. We also thank H.R. Andreasyan
for the help in the work.

This research has made extensive use of \textit{Aladin} sky atlas, \textit{VizieR}
catalogue access tool,  \textit{VizieR}
photometry tool and \textit{SIMBAD} database, which are developed and
operated at CDS, Strasbourg Observatory, France.  This work has made use of data from the European Space Agency (ESA) space mission Gaia, processed by the Gaia Data Processing and Analysis Consortium (DPAC). Funding for the DPAC is provided by national institutions, in particular the institutions participating in the Gaia MultiLateral Agreement (MLA). The Gaia mission website is \url{https://www.cosmos.esa.int/gaia}. The Gaia archive website is \url{https://archives.esac.esa.int/gaia}. This publication makes use of data products from the Two Micron All Sky Survey (2MASS), which is a joint project of the University of Massachusetts and the Infrared Processing and Analysis Center/California Institute of Technology, funded by the National Aeronautics and Space Administration and the National Science Foundation. This publication also makes use of data products from the Wide-field Infrared Survey Explorer (WISE), which is a joint project of the University of California, Los Angeles, and the Jet Propulsion Laboratory/California Institute of Technology, funded by the National Aeronautics and Space Administration. The images from Spitzer Enhanced Imaging Products program were used via NASA/IPAC
Infrared Science Archive \url{https://www.ipac.caltech.edu/doi/irsa/10.26131/IRSA3}

\section*{Data Availability}

The data underlying this article will be shared on reasonable request to the corresponding author.

\end{document}